\documentstyle[aps,multicol,epsfig]{revtex}
\newcommand {\be}{\begin{equation}}
\newcommand {\ee}{\end{equation}}
\newcommand {\bea}{\begin{eqnarray}}
\newcommand {\eea}{\end{eqnarray}}
\newcommand {\NAT}[1] {{\it Nature} {\bf{#1}}}
\newcommand {\PRE}[1] {{\it Phys. Rev. E} {\bf{#1}}}
\newcommand {\PRL}[1] {{\it Phys. Rev. Lett.} {\bf{#1}}}
\newcommand {\SCI}[1] {{\it Science} {\bf{#1}}}
\newcommand {\PYSCA}[1]  {{\it Physica A} {\bf{#1}}}
\newcommand {\JSP}[1] {{\it J. Stat. Phys.} {\bf{#1}}}
\newcommand {\EQ}[1] {{Eq. (\ref{#1})}}

\newcommand {\nonum}{\nonumber}

\begin{document}
\draft
\title{Weighted Evolving Networks}
\author{S.H. Yook, H. Jeong, A.-L. Barab\'asi}
\address{Department of Physics, University of Notre Dame, Notre Dame, IN 46556}
\author{Y. Tu}
\address{IBM, TJ Watson Research Center, POB 218, Yorktown Heights, NY 10598 USA }

\date{\today}
\maketitle

\thispagestyle{empty}

\begin{abstract}

Many biological, ecological and economic systems are
best described by weighted networks,
as the nodes interact with each other with varying strength.
However, most network models studied so far are binary, the link 
strength being either 0 or 1.
In this paper we introduce and investigate the scaling properties of a class of models
which assign weights to the links as the network evolves.
The combined numerical and analytical approach indicates
that asymptotically the total weight distribution converges to the
scaling behavior of the connectivity distribution,
but this convergence is hampered by strong logarithmic corrections.
\end{abstract}

\begin{multicols}{2}
%\narrowtext
%\newpage
%\section{Introduction}
With the increased availability of detailed topological data 
on complex networks describing communication, social
or biological systems emerges the need to develop tools
to understand in general terms the origin and evolution of these
complex evolving systems.
The recognition that real networks are fundamentally different
from the random models that dominated the mathematical literature
in the past forty years\cite{bo85,ER}
lead to a surge of activity in addressing the statistical
properties of these systems\cite{w99,ws98,np,ab99,dor99,red00,amaral,havlin}. 
Despite rapid advances in uncovering the topology of complex networks,
in one aspect most models are incomplete when compared with real systems:
current network models assume that all links are equivalent.
But in many fields it is well known that 
the interaction strengths can vary widely,
such variations being essential to the network's ability to carry on
its basic functions.
Sociologists have repeatedly argued 
about the importance of assigning strengths to social links,
finding that the weak links people have outside their close circle of friends
play a key role in keeping the social system together\cite{gra}.
Recently, Newman has showed that assigning weights to the
links between scientists allows for a better characterization of
the scientific collaboration web\cite{new00-c}.
Similarly, there is an ongoing discussion about the importance of
weak links between species in guaranteeing the stability of an ecosystem\cite{be99};
in addition viewing the economy as a complex network of companies,
the monetary value of the various business transactions linking them
is a key component of the systemic characterization.
Finally, many transportation networks nature designed,
ranging from cardiovascular to respiratory networks, 
have well defined weights or flow rates assigned to the links,
whose magnitude is intimately determined by the network's topology\cite{ban99}.
Despite the known importance of interaction strengths
in various well known systems,
there have been no attempts to model networks other than binary nets,
whose links have weights $0$ or $1$.

In this paper we take a first step in the direction of a systematic
study of networks with non-binary connectivities.
We introduce and investigate two models that
assign weights to new links as they are dynamically created,
providing a prototype of a weighted evolving network.
While we choose the simplest possible models,
in which the weights are driven by the network connectivity only,
numerical simulations indicate that the distribution of the total weight
scales differently from the total connectivity.
However, an analytical solution reveals that the different scaling behavior 
can be explained by
strong logarithmic correction, and asymptotically
the investigated weighted networks belong to the same universality class
as their unweighted counterparts. 

{\it Weighted Scale-Free (WSF) Model} :
Starting from a small number ($m_0$) of vertices,
at each time step we add a new node which links to $m$ existing nodes in the system.
The probability that a new node $j$ will connect to a existing node $i$ is
\bea
\label{pi}
\Pi_i={k_i \over {\sum_j k_j}},
\eea
where $k_i$ is the total number of links that the node $i$ has. In assigning
a weight to the newly established link $j\leftrightarrow i$,
we assume that the weight $w_{ji}(=w_{ij})$ is proportional to $k_i$,
i.e., more connected (and therefore more "powerful") nodes gain more weight. 
Also, one can assume that all new nodes
have fairly uniform total `resources' for linking
to other nodes in the system, we therefore
require that each new node has a fixed total weight,
i.e. we normalize $w_{ij}$ such that
the sum of the weights for the $m$ new links 
is $\sum_{\{i^\prime\}} w_{ji^\prime}=1$,
where $\{i^\prime\}$ represents a sum over the $m$ existing nodes to which
the new node $j$ is connected.
As a result of the two assumptions, each link $i \leftrightarrow j$ of the
newly added node $j$
is assigned a weight as
\be
\label{def-3}
w_{ji} ={ {k_{i}} \over {\sum_{\{i^\prime\}}  k_{i^\prime}}}.
\ee
%In general one , a constraint incorporated into 
%the normalization (\ref{def-3}).
%That is, a new company will have limited momentary resources for trade,
%or a new acquittance is always a weaker link than a long and ongoing friendship.
%In the absence of normalization, nodes that join the system at some latter time 
%could have unrealistically large link weights 
%if they connect to some nodes with very large $k$
%(which is rather frequent thanks to preferential attachment (\ref{pi})).

{\it Weighted Exponential (WE) Model} : 
The model is inspired by model A discussed
in Refs. \cite{ajb99,baj99}, and is defined as follows : 
at every time step we add a new node with $m(\le m_0)$ links,
connected with {\it equal probability}
to the nodes present in the system.
The weights of the links are assigned again by using (\ref{def-3}).

The difference between the WSF and WE models comes in
preferential attachment, which is  known to
fundamentally alter the topology\cite{dor99,red00,amaral,ajb99,baj99,kk00-c} :
The WSF model generates a scale-free network
whose connectivity distribution follows $P(k)\sim k^{-3}$,
while the network generated by
the WE model is exponential with the connectivity distribution
following $P(k)={e \over m}e^{-k/m}$.
Since the weights of the links are driven by the connectivity,
this difference is expected to lead to significant changes
in the distribution of the link strengths as well.

We start by investigating the weight distribution
of the two models.
As Fig 1 (a) and (b) shows, both the WE and the WSF models lead to a
peaked and  skewed weight distribution,
whose tails decay exponentially (or faster) for large $w_{ij}$.
The boundedness of $P(w_{ij})$ is due to the normalization condition,
which does not allow individual weights to be larger than 1.
Most important, however, we find that the distribution is stationary,
i.e. $P(w_{ij})$ is independent of time (and system size).

While the individual weights assigned to links, $w_{ij}$, are bounded,
we get a very different picture
when we study the total weight associated with a selected node.
In binary networks the node's importance is
characterized by the total number of links it has, $k_i$.
Similarly, in a weighted network the importance of a node $i$
can be measured by its total weight,
obtained by summing the weights of the links that connect to it,
$w_i=\sum_{\{j\}} w_{ij}$.

Due to the normalization condition (\ref{def-3}) a {\it new} node has $w_i=1$,
but $w_i$ increases in time every time 
when a subsequently added nodes link to $i$.
Since in both models the weights are determined by the
network connectivity,
we expect that $P(w)$ closely follows $P(k)$.
In contrast, the numerical results summarized in Fig. 2 indicate striking differences
between $P(k)$ and $P(w)$.
As Fig 2a shows, while for the WE model $P(k)$ decays exponentially,
$P(w)$ systematically deviates from a simple exponential behavior.
This difference is even more evident in the network dynamics:
while both $k_i(t)$ and $w_i(t)$
appear to increase logarithmically in time, they can be fitted with 
a different slope on a $\log$-linear plot (Fig. 2b).
Similar systematic discrepancies are observed for the WSF model as well:
as Fig. 2c indicates, while $P(w)$ can be fitted by a power law,
$P(w)\sim w^{-\sigma}$,
it appears that the exponent $\sigma$ is different from $\gamma =3$.
Furthermore, we find that $\sigma$ depends strongly on $m$ (Fig. 2c).
Again, this difference is reflected in the dynamical
behavior of $k_i(t)$ and $w_i(t)$: 
as Fig. 2e indicates, $w_i(t)\sim t^{\beta}$ with $\beta > 1/2$,
in contrast with $k_i(t)\sim t^{1/2}$\cite{ab99,baj99}
predicted by the binary scale-free model.

To understand the different behaviors
of $w_i$ and $k_i$ uncovered by the numerical simulations,
we resort to analytical method in determining the averaged behavior of 
$w_i(t)$ for the discussed model.
To simplify the discussion in the following we assume $m=2$, however,
the calculations can be generalized for arbitrary $m$.
%Since $P(w_{ij})$ is bounded and independent of time (Fig. 1),
%we can assume that every time a new node $j$ connects to node $i$,
%the old node $i$ increases its weight on average with
%$\left< w_{ij} \right>=\int P(w_{ij})w_{ij} dw_{ij}$.
The total weight of node $i$ at time $t$ can be written as
\bea
w_i(t) &=& 1+\sum_{\{j\}} w_{ij}
 = 1+\int_{t_i^0}^{t} \tilde P_i(t^{\prime}) \left< w_{ij}(t^{\prime})\right> 
 dt^{\prime} ,
\label{wit2}
\eea
where $\tilde P_i(t)$ is the probability 
that node $i$ is selected to be connected to a new node $j$ at time $t$
and $t_i^0$ is the time at which the node $i$ has been added to the system.
$\left< w_{ij} \right>$ is the average weight of link $i \leftrightarrow j$
once the link is established. 
When a new node $j$ and the list of $m$ nodes $\{i^\prime\}$
to which it connects are selected, the weights of the links, $w_{ji^\prime}$ 
are assigned according to  (\ref{def-3}).
These weights depend on the number of links
the selected nodes have, i.e. $\{k_{i^\prime}\}$.
If we assume that node $j$ is connected to nodes $i$ and $l$ ($m=2$),
we have
\be
\left< w_{ij}(t)\right>=\int_{m}^{\infty} w_{ji}(l) {\mathcal P}(k_l)~dk_l
\label{avewij}
\ee
where $w_{ji}(l)$ is the weight between the $j$ and $i$ nodes, 
${\mathcal P}(k_l)$ is the probability distribution of $k_l$, the total link number of 
node $l$.
Substituting (\ref{avewij}) into (\ref{wit2}), we obtain
\be
w_i(t)=1+\int_{t_i^0}^t \int_{m}^{\infty} \tilde P_i(t^{\prime}) 
w_{ji}(l){\mathcal P}(k_l)~dk_l~dt^{\prime}.
\label{wit3}
\ee

According to (\ref{def-3}) for $m=2$, the weight $w_{ji}(l)$ is given by
\be
w_{ji}(l)={{k_i}\over{k_i+k_l}},
\ee

thus Eq.(\ref{wit3}) becomes
\be
w_i(t)=1+\int_{t_i^0}^t \int_{m}^{\infty} \tilde P_i(t^{\prime}) {{k_i}\over{k_i+k_l}}
{\mathcal P}(k_l)~dk_l~dt^{\prime}.
\label{fwit}
\ee

\EQ{fwit} represents a general expression for calculating 
$w_i(t)$ for $m=2$.
To apply it to the WE and WSF models, we need to calculate explicitly
$\tilde P(t)$ and ${\mathcal P}(k_l)$.

{\it WE model}: In the WE model the nodes to which a new node connects to are
selected uniformly among all existing nodes, 
thus the probability that node $i$ will be picked is 
independent of this node's connectivity and is given by 
\bea
\label{m0}
&\tilde P_i(t)&={m \over {t+m_0}}.
\eea
Similarly, the connectivity distribution
and the dynamical behavior of a single node are given by\cite{baj99}
\bea
\label{wsfkt}
& {\mathcal P}(k)&=Ae^{-k/m}={e \over m}e^{-k/m},\\
\nonum
&k_i(t)&=m\left[\ln(m_0+t-1) -\ln (m_0+t_i^0-1)+1 \right]\\
\nonum
& &	=m\left[\ln(at+b)+1\right],
\eea
where $a={1 \over{m_0+t_i^0-1}}$, $b={{m_0-1}\over{m_0+t_i^0-1}}$ and
the normalization condition is $1=\int_{m}^{\infty} {\mathcal P}(k)dk$.

Substituting (\ref{wsfkt}) into (\ref{fwit}), we obtain
\bea
\nonum
w_i(t)&=& 1+e\int_{t_i^0}^t \int_{m}^{\infty}{1\over{t^{\prime}+m_0}} 
	{{k_i(t^{\prime})}\over{k_i(t^{\prime})+k_l}}  e^{-k_l / m}~dk_l~dt^{\prime}.
\eea
After performing the integration
and inserting $k_i(t)$ from (\ref{wsfkt}), for large $t$ we obtain
\be
w_i(t)\simeq m \ln (at+b)-m\ln(\ln(at+b)+2)+C,
\ee
where $C$ is an integration constant independent of $t$.
Therefore the relation between $w_i(t)$ and $k_i(t)$
for large $t$ follows
\be
\label{ln-cor}
w_i(t)\simeq k_i(t) -m\ln\ln t+C.
\ee
The prediction (\ref{ln-cor}) is fully supported by numerical simulations:
in Fig. 3a we plot the difference $w_i(t)-k_i(t)$ as function of $\ln\ln(t)$,
showing that the difference indeed follows a double logarithmic law.
This result is very interesting since it indicates that the different slopes
observed in Fig. 2b for $k_i(t)$ and $w_i(t)$ do not represent distinct
power law scaling behaviors, but are the result of logarithmic
corrections.

%%%%%%%%%%%%%%%%%%%%%%%%%%%%%%%%%%%%%%%%%%%%%%%%%%%%%%%%%%%%%%%%%%%%%%%%%%%%%
%%%%%%% For model-A
%%%%%%%%%%%%%%%%%%%%%%%%%%%%%%%%%%%%%%%%%%%%%%%%%%%%%%%%%%%%%%%%%%%%%%%%%%%%%
%%%%\newpage
{\it WSF model}:
In the scale-free model the probability distributions and $k_i(t)$ are
given by \cite{baj99}
\bea
\label{ma}
\nonum
&\tilde P_i(t)&=m{k_i(t) \over \sum_{j}^{t} k_j}=m {{k_i(t)}\over{2mt}}
	={k_i(t) \over {2t}},\\
& {\mathcal P}(k)&=mk^{-2}~~~~~(\propto k\cdot P(k)),\\
\nonum
&k_i(t)&={m \over {\sqrt{t_i^0}}}\sqrt{t}.
\eea
Substituting (\ref{ma}) into (\ref{fwit}), and performing the integrals 
we obtain
\bea
\label{ln-cor-2}
w_i(t)&\simeq& k_i(t)
						-{m\over 8}\left( \ln {{m^2 t}\over t_i^0}\right)^2 
						+{m \over 2}\ln m \ln {t\over t_i^0}+C^{\prime},
\eea
indicating that despite a different scaling behavior
suggested by the numerical simulations (Fig. 2e),
we are dealing with strong logarithmic corrections and 
asymptotically we have $\beta^\prime = \beta$.
Again, the analytical prediction (\ref{ln-cor-2}) is confirmed 
by more detailed numerical simulations shown in Fig. 3b.

Our ability to calculate analytically $w_{ij}$ for the discussed models
is based on the fact that
the weights are driven by the connectivity distribution.
To address the generality of our results we investigated several
extensions of these two models, 
that we discuss in the following.

{\it Weight driven weight-}
In general one could expect that in some systems
the quantity determining the weight is not the connectivity, 
but are the weights themselves.
To investigate this possibility we replaced (\ref{def-3}) with
\be
\label{def-1}
w_{ji}={w_i \over {\sum_{i^\prime} w_{i^\prime}}}.
\ee
i.e. the weight of the newly added links are determined 
by the total weight of the nodes.
While we cannot solve this model analytically, the numerical
results are similar to those observed for the WE and WSF models:
an apparently different scaling behavior for $k$ and $w$ can be attributed
to slow corrections to scaling.

{\it Weight driven connectivity-}
In some systems the topology
could be driven by the total weights, and not by the connectivity.
Thus we assume that the probability (\ref{pi})
that a new node is connected
to a node $j$ is
\bea
\label{determine-1}
\Pi_i = {{w_i}\over{\sum_j w_j}},
\eea
where $w_i$ is the weight of node $i$.
The weights are then assigned following (\ref{def-3}).
We find that the scaling of this network is identical to
that of the scale-free model, and the evolution of the weights also
follows the paradigm established for the WSF model.

{\it Discussion-} Weighted links are a common feature of real networks,
thus addressing their scaling behavior
is of primarily importance if we are to understand complex networks in general.
Here we take a first step in this direction by investigating the 
scaling properties of several simple models
that incorporate mechanisms to assign weights to the links.
The weight distribution in the discussed models can be
determined analytically and by using numerical simulations,
allowing for a thorough analysis of the model's scaling properties.
A comparison between the analytical and numerical results
brings deeper understanding of the observed scaling behavior for the distribution functions. 
%is not free of surprises, however.
First, extensive simulations of networks whose size is
comparable to the real networks that are
currently available indicate the emergence of new scaling exponents
for the behavior of the total weights.
However, the analytical solutions reveal that the results are affected
by strong logarithmic corrections, and asymptotically the scaling behaviors
of the weighted and unweighted models are identical.
This result raises important questions regarding our ability to uncover 
the correct scaling behavior of real weighted networks, should such data
become available in the near future: the real exponents could be easily shadowed by 
corrections to scaling similar to that encountered in the investigated models here.

An important feature of the studied models is the fact that the distribution
of the individual weights, $w_{ij}$, is bounded and stationary.
This implies that the scaling behavior of the total weight distribution can not 
be explained by the distribution of individual weights. Indeed it is the 
correlation between the 
total weight and the linkage probability of the new node that drives the distribution towards power 
law behavior.
In general, one can imagine systems and models where the link distribution, $P(w_{ij})$, 
could also have nontrivial
distribution, such as a power law.
Investigating such models could be of major future interest.

The results presented in this paper represent only
the starting point towards understanding weighted networks.
In some real systems, diverse dynamical rules can 
govern the assignment of weights to links, which could
result in statistical properties of the network that are different from that discussed here.
In particular, we assumed that once a weight has been assigned to a link,
it stays unchanged,
which is often not the case in more realistic networks: weights can evolve
dynamically just as the network topology does.
For example, acquaintance can turn into friendship by strengthening a
previously weak link.
Determining the generic behavior of such complex evolving systems is a real
challenge for future research.
Despite these limitations, the investigated models give a glimpse into the 
complex behavior we are facing as we attempt to make network modeling more
realistic by incorporating weights.
Finally, the ultimate understanding of 
weighted networks will be determined by the available data on real systems.
While currently such data is rare, we believe that the increasing
interest in network modeling and creative data collection methods will
soon lead to the development of such data sets, offering further
guidance for modeling these complex systems.

Research at Notre Dame supported by NSF, PHY-9988674 and CAREER DMR97-01998.

\begin{center}

\begin{figure}
\psfig{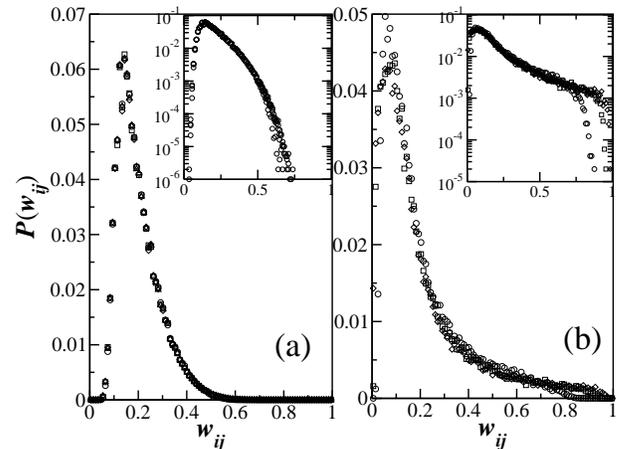}
\caption{
The distribution $P(w_{ij})$ of the individual link weights,$w_{ij}$ 
for the (a) WE and the (b) WSF models, defined in the text ($m=2$).
The symbols correspond to different system sizes (or time), i.e.
$N=10^3$($\bigcirc$), $10^4$($\Box$), $10^5$($\diamondsuit$) and
$10^6$($\triangle$).
The insets shows the same data on a $\log-$linear plot, indicating
that the tail decays faster than exponential.
}
\label{w-dist}
\end{figure}

\begin{figure}
\psfig{file=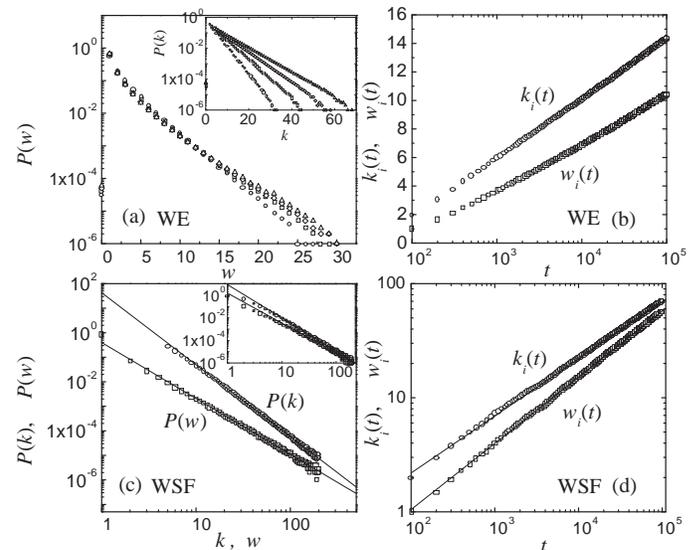,width=9cm}
%\centerline{\epsfxsize=9cm \epsfbox{fig2.eps}}
\caption{(a) Distribution $P(w)$ of the total connectivity $w$ assigned to 
individual nodes for the WE model. 
The symbols correspond to different values of $m$, i.e.
$m=2$ ($\bigcirc$), $3$ ($\Box$), $4$ ($\diamondsuit$) and $5$ ($\triangle$).
The inset shows the connectivity distribution, $P(k)$, for the same parameters
as in the main panel.
(b) Time dependence of $k_i(t)$ ($\bigcirc$)
and $w_i(t)$ ($\Box$) for a randomly selected node $i$
for the WE model ($i=5000$).
(c)$P(k)$ ($\bigcirc$) and $P(w)$ ($\Box$) distributions 
for the WSF model for $m=5$.
The inset shows the same data for $m=2$. 
(d) $k_i(t)$ ($\bigcirc$), $w_i(t)$ ($\Box$) vs. $t$ 
for the WSF model ($i=10000$).
}
\label{fig1}
\end{figure}

\begin{figure}
\psfig{file=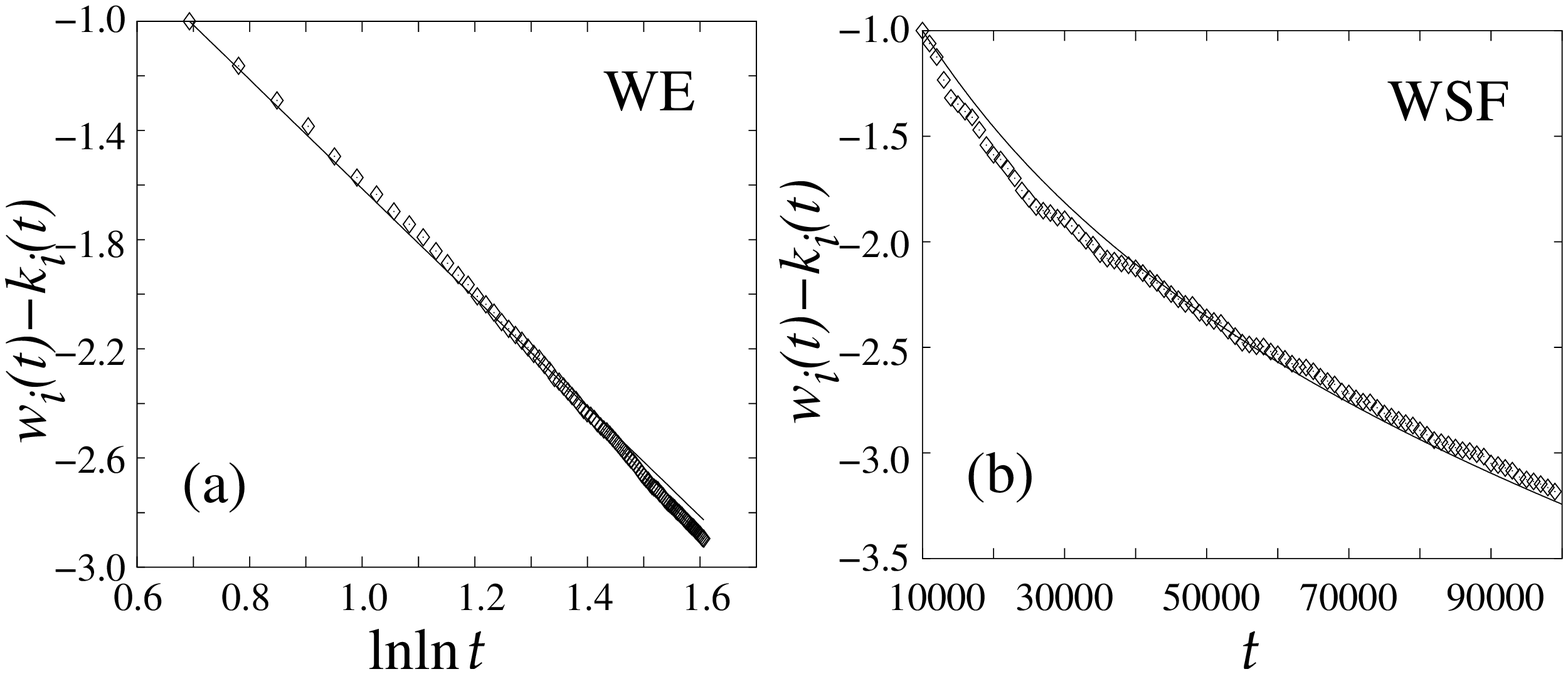,width=9cm}
\caption{The difference $(w_{i}(t)-k_i(t))$ for the
(a) WE and  the (b) WSF models.
The continuous lines in each case represent the analytic
solution (11) and (13), respectively.
We limited the simulations to nodes appears at large $t_i^0(t_i^0=10^4)$
to capture the asymptotic limit,
that is predicted by our predictions (11) and (13).
We find that for smaller $t_i^0$ the crossover time for 
the convergence to the analytic solution is numerically prohibited.}
\label{fig2}
\end{figure}

\end{center}

\end{multicols}

\end{document}